\documentclass[%
twocolumn,
superscriptaddress,
 amsmath,amssymb,
pra,
]{revtex4-1}

\usepackage{graphicx}% Include figure files
\usepackage{dcolumn}% Align table columns on decimal point
\usepackage{bm}% bold math
\usepackage[scaled]{helvet}
\usepackage{xcolor}

\begin{document}

%\preprint{APS/123-QED}

\title{Full orbital decomposition of Yu-Shiba-Rusinov states based on first principles}% Force line breaks with \\

\author{Tom G. Saunderson}
\email{tsaunder@uni-mainz.de}
\affiliation{HH Wills Physics Laboratory, University of Bristol, Tyndall Ave, BS8 1TL, United Kingdom}
\affiliation{Institute of Physics, Johannes Gutenberg University Mainz, 55099 Mainz, Germany}
\author{James F. Annett}%
\affiliation{HH Wills Physics Laboratory, University of Bristol, 
Tyndall Ave, BS8 1TL, United Kingdom}
\author{G\'abor Csire}%
\affiliation{Catalan Institute of Nanoscience and Nanotechnology (ICN2), CSIC, BIST, Campus UAB, Bellaterra, Barcelona, 08193, Spain}
\author{Martin Gradhand}%
\affiliation{HH Wills Physics Laboratory, University of Bristol, Tyndall Ave, BS8 1TL, United Kingdom}
\affiliation{Institute of Physics, Johannes Gutenberg University Mainz, 55099 Mainz, Germany}

\date{\today}% It is always \today, today,

\begin{abstract}
We have implemented the Bogoliubov-de Gennes (BdG) equation in a screened Korringa-Kohn-Rostoker (KKR) method for solving, self-consistently, the superconducting state for 3d crystals including substitutional impurities. In this report we extend this theoretical framework to allow for collinear magnetism and apply it to fcc Pb with 3d magnetic impurities. In the presence of magnetic impurities, there is a pair-breaking effect that results in sup-gap Yu-Shiba-Rusinov (YSR) states which we decompose into contributions from the individual orbital character. We determine the spatial extent of these impurity states, showing how the different orbital character affects the details of the YSR states within the superconducting gap. Our work highlights the importance of the first principles based description which captures the quantitative details making direct comparisons possible with experimental findings.
\end{abstract}

\pacs{Valid PACS appear here}% PACS, the Physics and Astronomy

\maketitle

\section{\label{sec:1}Introduction}

Magnetism and conventional superconductivity are phenomena related to competing types of order. Theoretically, it was first shown by Abrikosov and Gor'kov~\cite{Abrikosov1959} that at a paramagnetic impurity concentration of $\sim 1\%$ in a superconductor the spectral energy gap would no longer correspond to the ordering parameter $\Delta$. For higher concentrations of impurities the spectral gap can even vanish totally, leading to gapless superconductivity characterised by a finite T$_C$ and $\Delta$ but with no spectral gap.\cite{Phillips1963} Local, real space models were subsequently constructed by Yu \cite{Yu1965}, Shiba \cite{Shiba1968} and Rusinov \cite{Rusinov1969} using local, one band models around a classical impurity spin. They predicted the existence of a pair of localised, in-gap (YSR) states either side of the Fermi energy associated with the exchange splitting $J$ of the spin. 

Subsequently, experiments investigating YSR states have predicted multiple pairs of in-gap states \cite{Ji2008,Hatter2015,Ruby2016,Choi2017,Liebhaber2020}, where it was argued that the origin of multiple resonances may arise from magnetic anisotropy \cite{Hatter2015}, the orbital character \cite{Ruby2016,Choi2017} or modulations in the charge density \cite{Senkpiel2019,Liebhaber2020}. Ruby \textit{et al} \cite{Ruby2016} investigated the (001) surface of Pb with a Mn impurity adsorbed onto the surface. They argue that the multiple YSR resonances originate from the crystal field splitting of the Mn d-orbitals. Using energy considerations and real space $dI/dV$ maps they were able to assign the relevant orbitals to the YSR resonances.

When investigating more complex superconductors like NbSe$_2$ \cite{Senkpiel2019,Liebhaber2020} or $\beta-$Bi$_2$Pd \cite{Choi2018} it becomes immediately obvious that disentangling hybridised YSR peaks, or YSR peaks entangled with coherence peaks will become increasing challenging. Symmetry arguments and energy consideration will not be sufficient to uniquely assign the large number of in-gap resonances. In this report we build upon previous work \cite{Saunderson2020, Saunderson2020_a} to address this problem from first principles, considering all electrons and their magnetic orderings fully while capturing superconductivity within a one-parameter model of local BCS type pairing. 

We use this formalism to investigate the 3d series of elements as impurities in fcc Pb. After a brief introduction of the methodological implementations we show our results for all magnetic 3d impurities discussing the distinct YSR resonance pairs arising from the anticipated t$_{2g}$ and e$_{g}$ orbitals. However, beyond that we will highlight the existence of $l=0$ (s-electron) YSR resonances enforcing the necessity of an all-electron description. Finally, we will analyse the spatial decay of the magnetism as well as the in-gap states within the superconducting Pb, presenting the orbital-resolved local densities.   

\section{\label{sec:2}Method}

The implementation of the Bogoliubov-de Gennes (BdG) equation in the Korringa-Kohn-Rostoker Greens function method~\cite{Saunderson2020} has been described earlier as well as the extension to real space impurity systems~\cite{Saunderson2020_a}. Here, we present a further step namely the incorporation of collinear magnetism and in the following we will restrict the technical discussion to the equations relevant for this development.

In order to incorporate magnetism and superconductivity~\cite{Oliveira1988} into density functional theory~\cite{DFT1990}, three effective potentials are required, the electron potential $V_{eff}(\mathbf{r})$, the magnetic field $B_{eff}(\mathbf{r})$ and the effective pairing potential $\Delta_{eff}(\mathbf{r})$,
\begin{align}
\label{eqn:Veffspin}
V_{eff}(\mathbf{r}) &= V_{ext}(\mathbf{r}) + \int d^3 r  \frac{\rho(\mathbf{r})}{|\mathbf{r}-\mathbf{r}'|} + \frac{\delta E_{xc}[\rho,m]}{\delta \rho(\mathbf{r})}, \\
\label{eqn:meffspin}
B_{eff}(\mathbf{r}) &= B_{ext}(\mathbf{r}) + \frac{\delta E_{xc}[\rho,m]}{\delta m(\mathbf{r})}, \\
\label{eqn:Deltaeffspin}
\Delta_{eff}(\mathbf{r}) &= \Lambda\chi(\mathbf{r}).
\end{align}
Here, $E_{xc}[\rho,m]$ is the exchange correlation functional for
the normal state, $\rho(\mathbf{r})$, $m(\mathbf{r})$ are the usual
charge and spin densities and $\chi(\mathbf{r})$ is the  anomalous density. Finally, $\Lambda$ is the interaction parameter~\cite{Saunderson2020}, which is the one free parameter in our description typically fixed to recover experimentally observed gap sizes. This framework, using a simplified phenomenological parameter, was introduced in Ref.~ \cite{Suvasani1993} and subsequently implementation were presented in Refs.~\cite{Csire2015a,Csire2018}. It was already been shown to effectively describe gap anisotropy \cite{Saunderson2020} and impurity scattering \cite{Saunderson2020_a}, along with complex superconducting order parameters in LaNiC2~\cite{Csire2018a} and LaNiGa2~\cite{Ghosh2020}
Within the non-relativistic theory the densities are given by
\begin{align}
\label{eqn:rhoupdn}
\rho(\mathbf{r}) =& \rho_{\uparrow}(\mathbf{r}) + \rho_{\downarrow}(\mathbf{r}), \\
\label{eqn:mupdn}
m(\mathbf{r}) =& \rho_{\uparrow}(\mathbf{r}) - \rho_{\downarrow}(\mathbf{r}), \\
\label{eqn:chiupdndnup}
\chi_S(\mathbf{r}) =& \frac{1}{2}\left(\chi_{\uparrow\downarrow}(\mathbf{r}) - \chi_{\downarrow\uparrow}(\mathbf{r})\right).
\end{align}
Hence, the resulting spin BdG Hamiltonian is defined as,
\begin{equation}
\label{eqn:BdGspin}
\hat{H}_{BdG}(\mathbf{r}) =
\left(\begin{matrix}
\hat{H}^{\uparrow\uparrow}(\mathbf{r}) & 0 &
0 & \Delta^{\uparrow\downarrow}_{S}(\mathbf{r})\\
0 & \hat{H}^{\downarrow\downarrow}(\mathbf{r}) &
\Delta^{\downarrow\uparrow}_{S}(\mathbf{r}) & 0\\
0 & \Delta^{\downarrow\uparrow}_{S}(\mathbf{r})^* &
-\hat{H}^{\uparrow\uparrow}(\mathbf{r})^* & 0 \\
\Delta^{\uparrow\downarrow}_{S}(\mathbf{r})^* & 0 &
0 & -\hat{H}^{\downarrow\downarrow}(\mathbf{r})^* 
\end{matrix}\right)\ \textrm{,}
\end{equation}
where
\begin{align}
\hat{H}^{\sigma\sigma}(\mathbf{r}) =& \hat{H}_{0}(\mathbf{r}) + V^{\sigma\sigma}_{eff}(\mathbf{r}), \\
V^{\uparrow\uparrow}_{eff}(\mathbf{r}) =& V_{eff}(\mathbf{r}) + B_{eff}(\mathbf{r}), \\
V^{\downarrow\downarrow}_{eff}(\mathbf{r}) =& V_{eff}(\mathbf{r}) - B_{eff}(\mathbf{r}), \\
\Delta^{\uparrow\downarrow}_{S}(\mathbf{r}) =& +\Lambda \chi_{S}(\mathbf{r}), \\
\Delta^{\downarrow\uparrow}_{S}(\mathbf{r}) =& -\Lambda\chi_{S}(\mathbf{r}).
\end{align}
Equation (\ref{eqn:BdGspin}) can be brought into a block-diagonal form  such that
\begin{equation}
\hat{H}^{\sigma\sigma^*}_{BdG}(\mathbf{r}) =
\left(\begin{matrix}
\hat{H}^{\sigma\sigma}(\mathbf{r}) & \Delta^{\sigma\sigma^*}_{S}(\mathbf{r})\\
\Delta^{\sigma^*\sigma}_{S}(\mathbf{r})^* & -\hat{H}^{\sigma^*\sigma^*}(\mathbf{r})^* 
\end{matrix}\right)\ \textrm{,}
\label{eqn:BdGspin_diag}
\end{equation}
where $\sigma=\{\uparrow,\downarrow\}$ and $\sigma^*$ represents the opposing spin to $\sigma$. The corresponding Green's function is defined as 
\begin{equation}
\hat{G}_{BdG}(z) = (z\hat{I} - \hat{H}_{BdG})^{-1}\ \textrm{,}
\end{equation}
where $\hat{H}_{BdG}(\mathbf{r}) = \langle \mathbf{r}|\hat{H}_{BdG}|\mathbf{r}\rangle$, and can be simplified into a block diagonal form accordingly
\begin{equation}
{G}_{BdG,\sigma}(z,\mathbf{r},\mathbf{r}') = \left(\begin{matrix}
G^{ee}_{\sigma\sigma}(z,\mathbf{r},\mathbf{r}') & G^{eh}_{\sigma\sigma^*}(z,\mathbf{r},\mathbf{r}') \\
G^{he}_{\sigma^*\sigma}(z,\mathbf{r},\mathbf{r}') &
G^{hh}_{\sigma^*\sigma^*}(z,\mathbf{r},\mathbf{r}')
\end{matrix}\right).
\end{equation}
The relevant densities are expressed by the Green's function
\begin{align}
\label{eqn:rho}
\rho_\sigma(\mathbf{r}) = &-\frac{1}{\pi}\int^{\infty}_{-\infty} d\epsilon f(\epsilon) \mathrm{Im}\mathrm{Tr}G^{ee}_{\sigma\sigma}(\epsilon,\mathbf{r},\mathbf{r}') \nonumber\\
&-\frac{1}{\pi}\int^{\infty}_{-\infty} d\epsilon [1-f(\epsilon)] \mathrm{Im}\mathrm{Tr}G^{hh}_{\sigma\sigma}(\epsilon,\mathbf{r},\mathbf{r}'),\\
\label{eqn:chi}
\chi_{\sigma\sigma^*}(\mathbf{r}) = &-\frac{1}{4\pi}\int^{\infty}_{-\infty} d\epsilon [1-2f(\epsilon)] \mathrm{Im} \mathrm{Tr} G^{eh}_{\sigma\sigma^*} (\epsilon,\mathbf{r},\mathbf{r}') \nonumber \\
&-\frac{1}{4\pi}\int^{\infty}_{-\infty} d\epsilon [1-2f(\epsilon)]  \mathrm{Im} \mathrm{Tr} G^{he}_{\sigma\sigma^*} (\epsilon,\mathbf{r},\mathbf{r}').
\end{align}

\begin{figure}[t]
\includegraphics*[width=1\linewidth,clip]{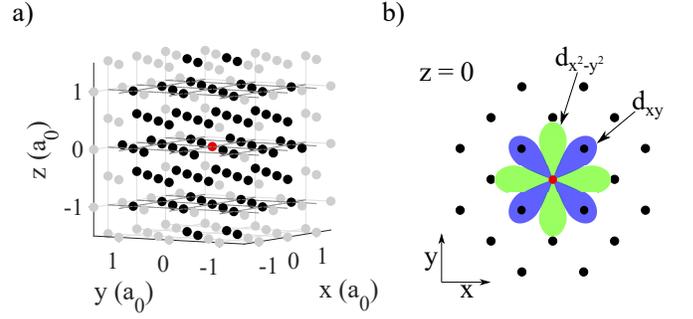}
\centering
\caption{Panel (a) shows the atomic sites around the impurity atom in units of the atomic spacing, $a_0=4.95\mathrm{\AA}$. Black dots represent atomic sites within the cluster, grey dots represent the unperturbed atomic sites outside the cluster, the red dot represents the impurity site. Panel (b) shows the $z=0$ crystal plane and the orientation of the $d_{x^2-y^2}$ (green) and $d_{xy}$ (blue) orbitals. } 
\label{fig:CLUSTER}
\end{figure}

Implementing this extension in the corresponding bulk \cite{Saunderson2020} and real-space impurity code \cite{Saunderson2020_a} will enable us to address the coupling between the superconducting state and magnetism. Here, we will focus on the effect of magnetic impurities with the corresponding in-gap YSR states. As test scenario we consider superconducting bulk fcc Pb  \cite{Saunderson2020} with the interaction parameter $\Lambda$ tuned such to the experimental gap size from Ruby~\textit{et al.}~\cite{Ruby2015b}. For the real-space impurity cluster embedded in the infinite periodic crystals we used 87 atoms as shown in panel Fig.~\ref{fig:CLUSTER}~(a). 

\section{\label{sec:3}Normal State Analysis}

\begin{figure}[b]
\includegraphics*[width=1\linewidth,clip]{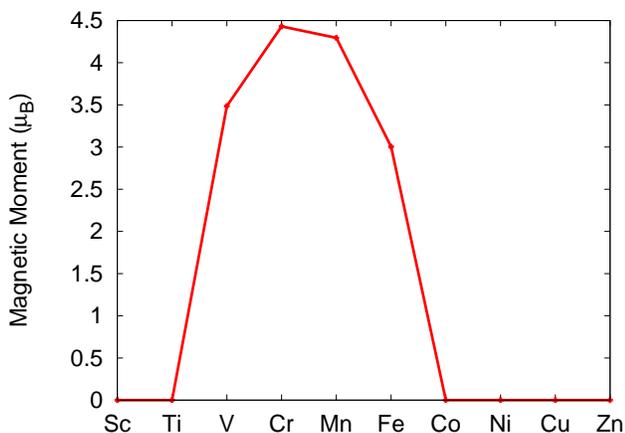}
\caption{The local magnetic moment for the 3d impurity embedded in a cluster of 87 Pb atoms.} 
\label{fig:MagImp}
\end{figure}

\begin{figure}[h]
\includegraphics*[width=1\linewidth,clip]{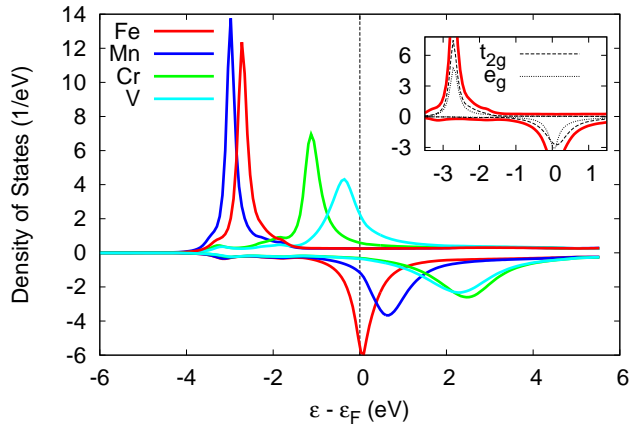}
\caption{The spin-resolved LDOS of all magnetic impurities embedded in normal state Pb. The inset shows the splitting between the e$_g$ and t$_{2g}$ orbitals for Fe.} 
\label{fig:NonSupMagImps}
\end{figure}

Starting with self-consistent scalar relativistic (SRA) solutions for the normal state, the resulting magnetic moments of the localised 3d impurity atoms are summarized in Fig.~\ref{fig:MagImp}. For the Pb host we identify only V, Cr, Mn and Fe to be magnetic. For all other 3d elements only the non-magnetic solutions were established self-consistently. The resulting local density of states (LDOS) for all magnetic impurities is shown in Fig.~\ref{fig:NonSupMagImps}. Notably, while for V it is the majority spin channel which is close to the Fermi energy, this impurity level shifts further away from the Fermi energy as we go through the 3d series. For the Fe impurity it is the minority spin channel which is situated right at the Fermi energy. For the elements in between we see a gradual transition between the two cases.

In the cubic fcc lattice the crystal field lifts the degeneracy of the d-orbitals and the impurity levels are split into the e$_g$ and t$_{2g}$ orbitals.~\cite{Dresselhaus2007} As an example this is shown for the Fe minority level as inset in Fig.~\ref{fig:NonSupMagImps}.

\begin{figure}[h]
\includegraphics*[width=1\linewidth,clip]{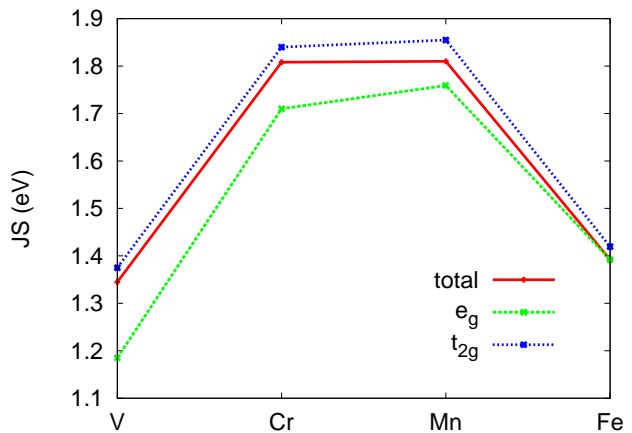}
\caption{The exchange energy, JS, for each magnetic impurity resolved for the total (red), the e$_{g}$ (green) and the t$_{2g}$ (blue) states.} 
\label{fig:JS}
\end{figure}

Within simplified models~\cite{Yu1965,Shiba1968,Rusinov1969,Balatsky2006} it has been shown that it is the energy splitting between the minority and majority levels which will determine the energy positioning of the in-gap states. For these models the impurity spin is considered as classical magnetic moment $S$ coupled to the electrons of the crystal via the exchange interaction $J$. The resulting energies are than expressed in terms of a T-matrix scattering approach. As the e$_g$ and t$_{2g}$ experience slightly different splitting this will result in a clear lifting of the degeneracy between those orbitals for the in-gap states. We have summarized the corresponding exchange energies
\begin{equation}
JS = \frac{p^{\uparrow}-p^{\downarrow}}{2}\ \text{,}
\end{equation}
in Fig.~\ref{fig:JS}. In the above definition $p^{\uparrow}$ and $p^\downarrow$ are the spin-up and spin-down peak positions, respectively. As expected, the trend follows the total magnetic moment being largest for Cr and Mn. Importantly, for V we find the strongest difference between the splitting in the  e$_g$ and t$_{2g}$ states which would suggest that the corresponding in-gap states will be well separated in energy. In contrast, this is weakest for Fe as indicated by the smallest separation of the in gap states.

\section{\label{sec:4}Superconducting State Analysis}

\begin{figure*}[t]
\includegraphics*[width=1\linewidth,clip]{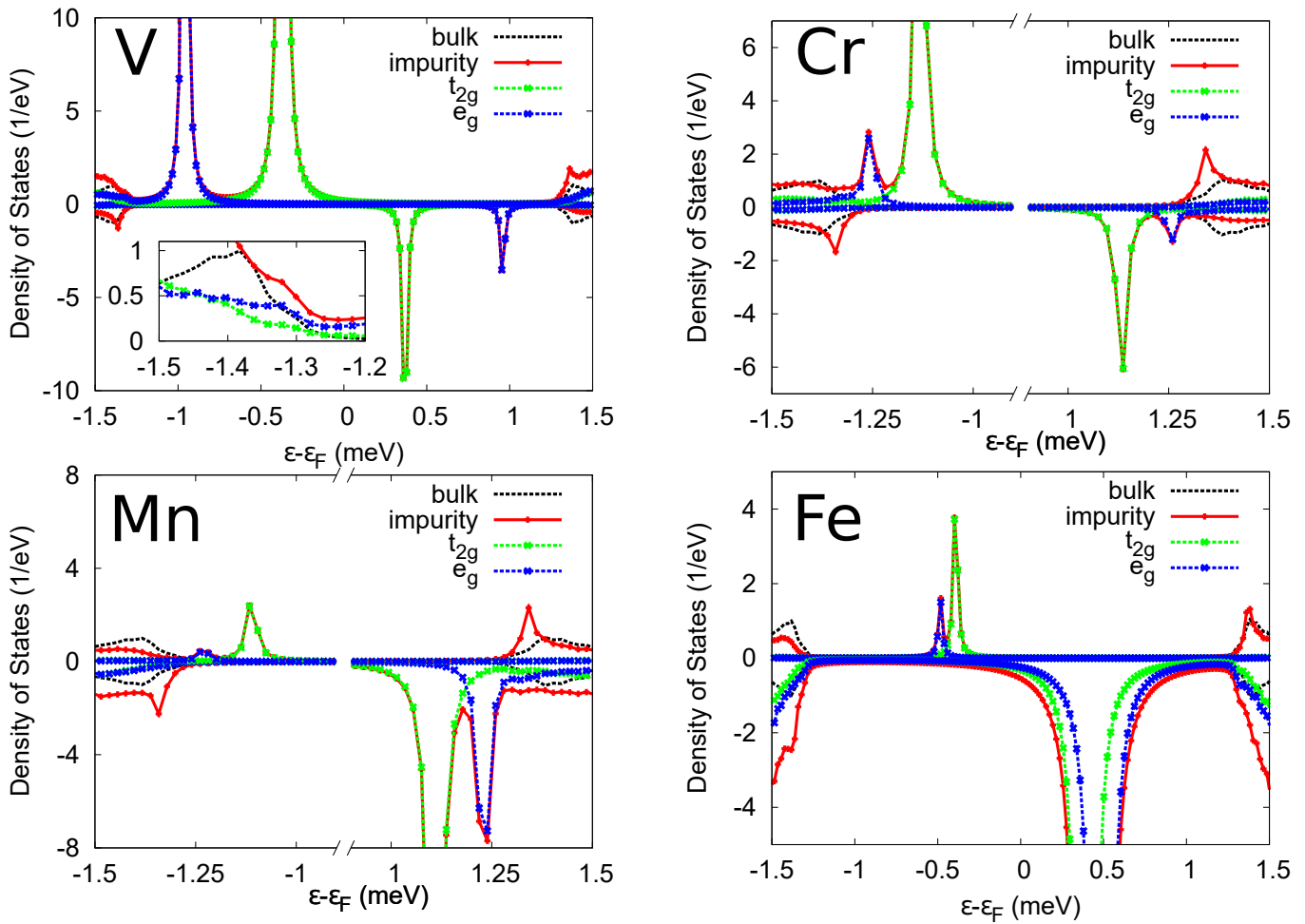}
\centering
\caption{The spin-resolved electronic LDOS (red) for the magnetic impurities (V, Fe, Mn, Cr), with the t$_{2g}$ (green dashed) and e$_{g}$ (blue dashed) densities and the bulk Pb DOS (black dashed) shown within the energy resolution of the superconducting gap.} 
\label{fig:MnImpDorbs}
\end{figure*}

Extending the analysis including superconductivity gives access to the YSR in-gap states. For each step we perform the corresponding normal state calculations first and extend it to the superconducting case in the following. In all cases we perform fully self-consistent SRA calculations based on the BdG Hamiltonian Eq.~\ref{eqn:BdGspin} to finally generate the electronic and magnetic structure for the magnetic atoms embedded in superconducting Pb. In all cases we set the interaction parameter $\Lambda$ at the impurity site to zero and keep it at the bulk value for all other sites.~\cite{Saunderson2020,Saunderson2020_a} The results are summarized in Fig.~\ref{fig:MnImpDorbs} for all magnetic impurities. In all four cases we find pronounced in gap states, well separated into e$_{g}$ and t$_{2g}$ states. As expected from the previous analysis the splitting between the e$_{g}$ and t$_{2g}$ is largest for V and hard to resolve for Fe. Table~\ref{table:BoundEnergies} summarizes the associated energies of the bound states. A clear correlation between Table~\ref{table:BoundEnergies} and the normal state exchange energies, summarized in Fig.~\ref{fig:JS},  is evident.

\begin{figure}[h]
\includegraphics*[width=1\linewidth,clip]{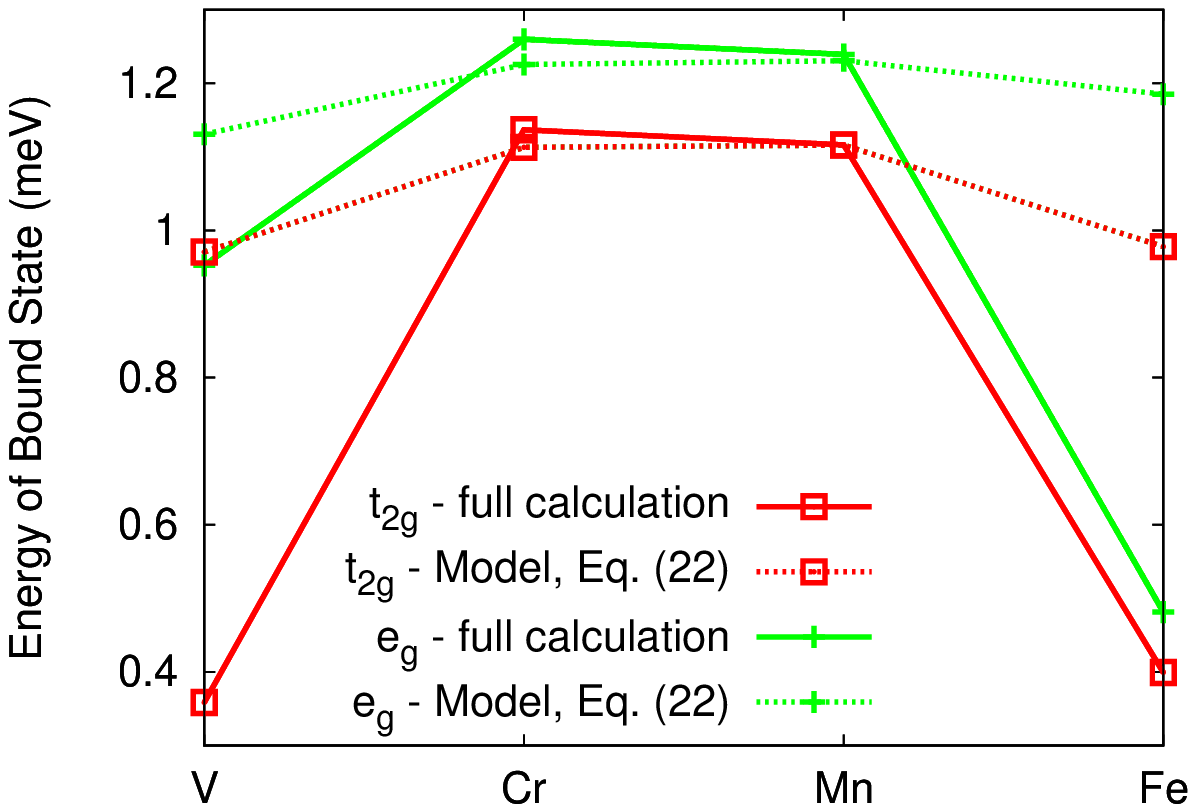}
\caption{Comparison of the energetic positions of the gap states as derived from the fully self-consistent calculations (Fig.~\ref{fig:MnImpDorbs}) to the simplified model (Eq.~\ref{eqn:YSRState}) of Refs.~ \onlinecite{Yu1965,Shiba1968,Rusinov1969,Balatsky2006}.} 
\label{fig:YSR_PEAKS}
\end{figure}

Furthermore, the superconducting symmetry implies that for each in-gap state there is a minority and a majority state symmetrically placed relative to the Fermi energy, which we indicated via the $\pm$ in Table~\ref{table:BoundEnergies}. While the energetic positions are forced to be symmetric the height of the corresponding states is determined by the normal state LDOS at the Fermi energy. As the minority state of the Fe impurity is perfectly placed at the Fermi energy (see Fig.~\ref{fig:NonSupMagImps}) this leads to the largest in-gap peak for the minority in-gap state (see Fig.~\ref{fig:YSR_PEAKS}). Similarly for V the majority level is closest to the Fermi energy leading to the corresponding in-gap state to be significantly larger than the minority peak. In contrast, for Cr both levels have similar distance to the Fermi energy in the normal state resulting in a very similar height for the Cr in-gap states. 

\begin{table}
\begin{tabular}{|c|c|c|c|c|}
\hline
\rule{0pt}{2.5ex} Impurity atom & V & Cr & Mn & Fe \\
\hline
\rule{0pt}{2.5ex} t$_{2g}$ state (meV) & $\pm 0.36$ & $\pm 1.13$  & $\pm 1.12$ & $\pm 0.40$ \\

\rule{0pt}{2.5ex} e$_g$ state (meV) & $\pm 0.95$& $\pm 1.26$  & $\pm 1.24$ & $\pm 0.48$  \\

\hline
\end{tabular}
\caption{Energetic positions of the in-gap bound states for the magnetic impurities in superconducting Pb.} 
\label{table:BoundEnergies}
\end{table}

In order to make the connection between the normal state impurity levels and the superconducting in-gap states more quantitative we follow previous models established for the YSR states~\cite{Yu1965,Shiba1968,Rusinov1969,Balatsky2006}. For isotropic (l=0) scattering the energy can be approximated to
\begin{equation}
\epsilon = \pm \Delta_0 \frac{1-\alpha^2+\beta^2}{\sqrt{(1-\alpha^2+\beta^2)^2+4\alpha^2}}\ \textrm{,}
\label{eqn:YSRState}
\end{equation}
where 
\begin{align}
\label{eqn:alphaYSR}
    \alpha = \pi N_0 JS\ \textrm{,} \\
\label{eqn:betaYSR}
    \beta = \pi N_0 V\ \textrm{.}
\end{align}  
Here, $\Delta_0$ is the bulk quasiparticle gap, $N_{0}$ is the density of states of Pb at the Fermi level in the normal state, and V is the non-magnetic scattering potential. Ignoring the correction from the scalar potential, $\beta=0$, but generalizing to the case where $N_0$ is different for e$_g$ and t$_{2g}$ states, the equation reduces to
\begin{equation}
\epsilon^{b}_{a} = \pm \Delta_0 \frac{1-\left(\alpha^{b}_{a}\right)^2}{1+{\left(\alpha^{b}_{a}\right)^2}}\ \text{,}
\label{eqn:YSR_moreSpecific}
\end{equation}
where $a$ stands for either e$_{g}$ and t$_{2g}$ states and $b$ is the index for the impurity V, Cr, Mn, Fe. Correspondingly, $\alpha^{b}_{a}$ generalises to
\begin{equation}
\alpha^{b}_{a} = \pi N^{a}_{0} (JS)^b_{a}\ \textrm{.}
\end{equation}
As the energies depend crucially on $N^{a}_{0}$ we decided to determine this parameter by fitting Eq.~\ref{eqn:YSR_moreSpecific} to the full {\it ab initio} calculations in the case of Mn. Here, we fixed $(JS)^{Mn}_{a}$ to the results shown in Fig.~\ref{fig:JS}. The resulting values are $N^{e_g}_{0}=0.966~({\mathrm eV})^{-1}$ and  $N^{t_{2g}}_{0}=0.593~({\mathrm eV})^{-1}$. In the following we kept $N^{a}_{0}$ the same for all impurities as it should capture the properties of the nonmagnetic host Pb and calculated the in-gap states using $(JS)^{b}_{a}$ for all the other impurities. The results in comparison to the directly extracted energies are shown in Fig.~\ref{fig:YSR_PEAKS}. While the agreement is far from perfect the trends are correctly reproduced. However, especially  for V and Fe, the elements with the largest and smallest splitting between the t$_{2g}$ and e$_g$ states, respectively, the model fails to reproduce the quantitative results of the full calculations. This highlights the importance of the full {\it ab initio} description as the model fails to capture the quantitative details making direct comparisons to experimental findings difficult.

A further conventionally made approximation is the restriction to $d$-orbitals only.~\cite{Ruby2016,Schrieffer1967} Given that the magnetism is dominated by the $d$ electrons and it is the magnetism which induces the in-gap states it appears natural to follow this approximation. However, in any real material all orbitals will hybridize and magnetism arising in the $l=2$ orbitals will ultimately induce spin-polarization in all other orbitals as well. Within our calculations this is particularly visible for Cr and Mn for which additional YSR peaks are visible near the Pb coherence peak (see Fig.~\ref{fig:MnImpDorbs}) at energies just above $(\epsilon-\epsilon_F)\sim1.3meV$. In this particular case it turns out they are $l=0$ orbital contributions. This finding highlights the fact that while in a first approximation it appears natural to reduce the discussion to the d orbitals, in details especially near the edge of the superconducting gap other orbitals might play a dominant role. Furthermore, determining which impurity has a large response from other orbitals will not be easy to predict without all-electron calculations.

\begin{figure}[h]
\includegraphics[width=1\linewidth,clip]{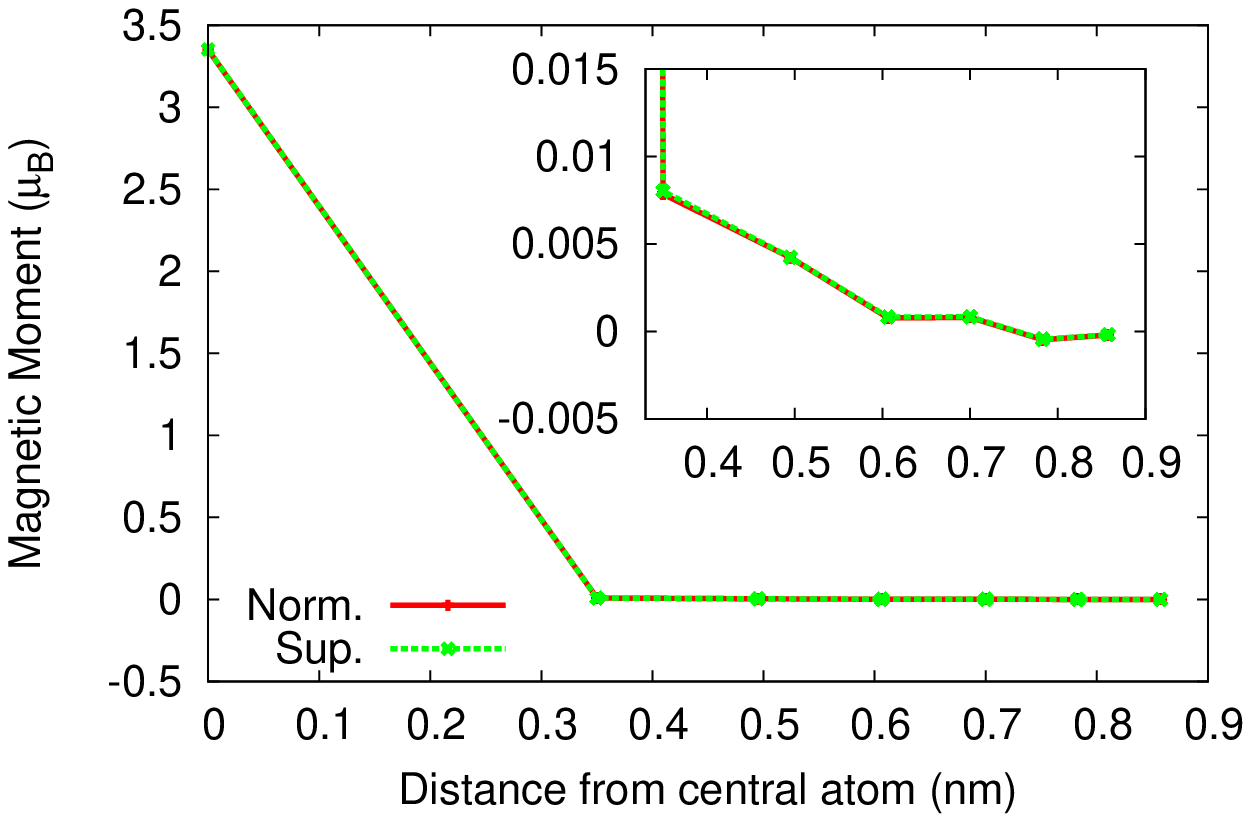}
\caption{Radial decay of the magnetic moment induced by a V impurity in the normal (green) and superconducting (red) state. The inset shows the decay beyond the nearest neighbour shell on a smaller scale.
} 
\label{fig:spatial_decay}
\end{figure}

\begin{figure*}[h]
\includegraphics*[width=1\linewidth,clip]{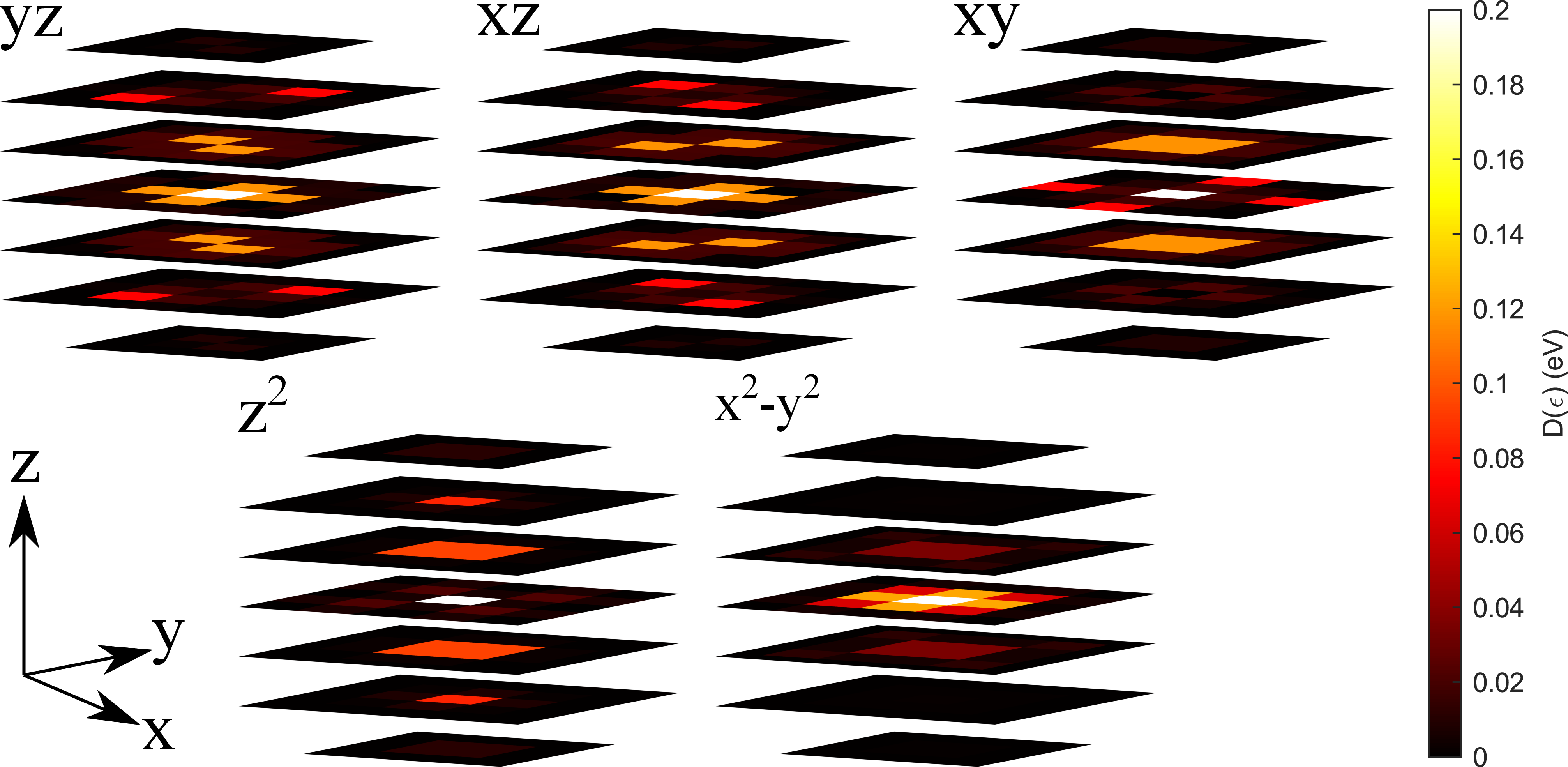}
\caption{Atom-resolved Charge densities at an energy corresponding to the minority in-gape state for the V impurity with the energy of $\epsilon = 0.36$meV ($d_{yz}$, $d_{xz}$ and $d_{xy}$ ) and  $\epsilon = 0.95$meV ($d_{z^2}$ and $d_{x^2-y^2}$). We show the z=\{-1.50,-1.0,-0.5,0.0,0.5,1.0,1.5\}$a_0$ planes around the impurity with a maximum in plane radius of $0.78$~nm. The colour scale is constraint so that the low values are amplified and the higher peak values are saturated.} 
\label{fig:Planar}
\end{figure*}
Finally, we would like to analyse the spatial dependence of the in gap states. In Fig.~\ref{fig:spatial_decay} the radial decay of the magnetic moment is shown for the case of a V impurity. It drops quickly to almost zero in the first shell already and oscillates weakly up to the 6th shell. For the total magnetic moment, there is no visible difference between the normal and the superconducting state in the spatial decay. In order to visualize the behaviour of the in-gap states we present the atom-resolved charge densities at the energy associated with the in-gap state in Fig.~\ref{fig:Planar}. In the case of the V impurities these energies are for the e$_g$ states $\epsilon=0.95$meV and for the t$_{2g}$ states $\epsilon = 0.35$meV. As discussed earlier, the cubic lattice leads to a splitting into e$_g$ and t$_{2g}$ states but does not lift the degeneracy of the $d_{z^2}$ and $d_{x^2-y^2}$ orbitals within the e$_g$-level nor for the $d_{yz}$, $d_{xz}$ and $d_{xy}$ orbitals within the t$_{2g}$ level. For this reason any visualisation for the two levels would preserve the cubic symmetry. Resolving all orbitals separately (see Fig.~\ref{fig:Planar}) highlights the power of full {\it ab initio} calculations. While the spatial resolution is limited to atomic sites the orbital characters are nevertheless clearly visible. The $d_{yz}$ and $d_{xz}$ reduce to a two-fold rotational symmetry around the z-axis rotated by $90$ degrees relative to each other. The other three orbitals show the corresponding four-fold rotational symmetry around the z-axis. Furthermore, the larger spatial extension of the $d_{z^2}$ orbital in the $z$ direction is clearly visible. Finally, as the the $d_{xy}$ and $d_{x^2-y^2}$ orbitals are rotated by $45$ degrees relative to each other this results in the $d_{x^2-y^2}$ pointing along the nearest neighbour bonds in the $z$ plane of the impurity atom. In contrast, the  $d_{xy}$ orbitals show the largest contribution out of plane as there the nearest neighbours are in direction of the orbital lobes. At a surface the degeneracy of these orbitals would be lifted and the orbital induced angular dependence of the density can be resolved with STM experiments.~\cite{Ruby2016,Choi2017,Ruby2018,Kuster2021}

\section{Summary}

We have extended previous work \cite{Saunderson2020,Saunderson2020_a} of implementing the BdG equations into the KKR method with substitutional impurities and collinear magnetism. As a model system we consider 3d impurities in a Pb three dimensional crystal inspired by experiments performed by Ruby~\textit{et al}~\cite{Ruby2016} investigating magnetism induced in-gap states at a superconducting surface. As predicted by simple models, the position and height of the induced in-gap states is strongly related to the normal state exchange splitting as observed for the magnetic impurities. However, while the over all correlation is clearly visible in the details the simplified model fails to make quantitatively correct predictions for the varying 3d impurity atoms. The most significant limitations of the model are the restriction to isotropic scattering and ignoring any hybridization of the d-electrons to other orbitals. Naturally, these models can be extended to more complex scenarios however rendering them more intractable as the parameter space increases. Within our real space superconducting DFT descriptions we are able to capture all orbital induced variations in their full complexity.

Beyond the consideration of the d-orbitals we established clear signatures of $l=0$ in-gap states induced via s-d hybridization. While these states are in close vicinity to the superconducting coherence peaks and as such difficult to observe in experiment, their existence highlights the complexity of any quantitative interpretation of experimental observation. Within our calculations these $l=0$ in-gap states were particular pronounced for Cr and Mn but hardly visible for V and Fe.  This again points to the importance of a correct all-electron description of the underlying band structure. This is especially relevant as similar YSR resonances have already been investigated \cite{Senkpiel2019}.

Finally, we investigate the radial decay of the magnetism and in-gap states inside the superconducting crystal. Interestingly, the decay of the magnetic moment is largely unaffected by the superconducting state. This is related to the fact that even in the normal state the magnetic moment decays quickly for the nearest neighbour atoms already. Nevertheless, by investigating the decay of the in-gap states for each orbital separately it was clearly possible to resolve the distinct orbital symmetries. This final step will enable us to make direct contact to STM experiments~\cite{Ruby2016} where the breaking of further spatial symmetries at the surface of the material will lift the degeneracies among the e$_{g}$ and t$_{2g}$ states.

In summary, we have performed fully self consistent calculations for magnetic impurities within the superconducting state using the BdG equations implemented within the KKR formalism. We have discussed qualitatively and quantitatively the formation of YSR resonances associated with the magnetic moment from the e$_{g}$ and t$_{2g}$ orbitals according to the cubic symmetry. Furthermore, we have established the existence of $l=0$ YSR resonances, highlighting the need for an all-electron description for even a qualitatively correct description of real materials. This is further strengthened by the fact that the position of the in-gap states is at best described qualitatively with simple models even for the dominant e$_{g}$ and t$_{2g}$ resonances. With the introduction of more impurities we could investigate the hybridisation of the YSR resonances resulting in the formation of YSR bands.~\cite{Zhang2020,Ruby2017} Similarly, the use of the KKR framework will enable us to extend the method to concentrated alloys exploiting the coherent potential approximation (CPA) as outlined previously.~\cite{Martin1999,Moradian2000,Moradian2000a} Finally, the incorporation of the fully relativistic BdG equations \cite{Csire2018} including spin-orbit coupling will, in a next step, enable us to investigate the existence of Majorana zero modes~\cite{Nadj-Perge2013,TopoSup2015,Kim2018,Schneider2020} at the surface of conventional superconductors.

\section{Acknowledgements}
This work was carried out using the computational facilities of the Advanced Computing Research Centre, University of Bristol - http://www.bris.ac.uk/acrc/. The above work was supported by the Centre for Doctoral Training in Condensed Matter Physics, funded by EPSRC Grant No. EP/L015544/1. G.C. gratefully acknowledges support from the European Union’s Horizon 2020 research and innovation program under the Marie Sklodowska-Curie Grant Agreement No. 754510. This work was supported by Spanish MINECO (the Severo Ochoa Centers of Excellence Program under Grant No. SEV- 2017-0706), Spanish MICIU, AEI and EU FEDER (Grant No. PGC2018-096955-B-C43), and Generalitat de Catalunya (Grant No. 2017SGR1506 and the CERCA Program). The work was also supported by the European Union MaX Center of Excellence (EU-H2020 Grant No. 824143). M.G. thanks the visiting professorship program of the Centre for Dynamics and Topology ad Johannes Gutenberg-University Mainz. The authors would like to thank M.-H. Wu and R. Gupta for many helpful discussions.


\begin{thebibliography}{36}%
\makeatletter
\providecommand \@ifxundefined [1]{%
 \@ifx{#1\undefined}
}%
\providecommand \@ifnum [1]{%
 \ifnum #1\expandafter \@firstoftwo
 \else \expandafter \@secondoftwo
 \fi
}%
\providecommand \@ifx [1]{%
 \ifx #1\expandafter \@firstoftwo
 \else \expandafter \@secondoftwo
 \fi
}%
\providecommand \natexlab [1]{#1}%
\providecommand \enquote  [1]{``#1''}%
\providecommand \bibnamefont  [1]{#1}%
\providecommand \bibfnamefont [1]{#1}%
\providecommand \citenamefont [1]{#1}%
\providecommand \href@noop [0]{\@secondoftwo}%
\providecommand \href [0]{\begingroup \@sanitize@url \@href}%
\providecommand \@href[1]{\@@startlink{#1}\@@href}%
\providecommand \@@href[1]{\endgroup#1\@@endlink}%
\providecommand \@sanitize@url [0]{\catcode `\\12\catcode `\$12\catcode
  `\&12\catcode `\#12\catcode `\^12\catcode `\_12\catcode `\%12\relax}%
\providecommand \@@startlink[1]{}%
\providecommand \@@endlink[0]{}%
\providecommand \url  [0]{\begingroup\@sanitize@url \@url }%
\providecommand \@url [1]{\endgroup\@href {#1}{\urlprefix }}%
\providecommand \urlprefix  [0]{URL }%
\providecommand \Eprint [0]{\href }%
\providecommand \doibase [0]{http://dx.doi.org/}%
\providecommand \selectlanguage [0]{\@gobble}%
\providecommand \bibinfo  [0]{\@secondoftwo}%
\providecommand \bibfield  [0]{\@secondoftwo}%
\providecommand \translation [1]{[#1]}%
\providecommand \BibitemOpen [0]{}%
\providecommand \bibitemStop [0]{}%
\providecommand \bibitemNoStop [0]{.\EOS\space}%
\providecommand \EOS [0]{\spacefactor3000\relax}%
\providecommand \BibitemShut  [1]{\csname bibitem#1\endcsname}%
\let\auto@bib@innerbib\@empty
%</preamble>
\bibitem [{\citenamefont {Abrikosov}\ and\ \citenamefont {{Gor'
  Kov}}(1959)}]{Abrikosov1959}%
  \BibitemOpen
  \bibfield  {author} {\bibinfo {author} {\bibfnamefont {A.~A.}\ \bibnamefont
  {Abrikosov}}\ and\ \bibinfo {author} {\bibfnamefont {L.~P.}\ \bibnamefont
  {{Gor' Kov}}},\ }\href {\doibase 10.1142/9789814366960} {\bibfield  {journal}
  {\bibinfo  {journal} {Sov. Phys. JETP}\ }\textbf {\bibinfo {volume} {8}},\
  \bibinfo {pages} {220} (\bibinfo {year} {1959})}\BibitemShut {NoStop}%
\bibitem [{\citenamefont {Phillips}(1963)}]{Phillips1963}%
  \BibitemOpen
  \bibfield  {author} {\bibinfo {author} {\bibfnamefont {J.~C.}\ \bibnamefont
  {Phillips}},\ }\href@noop {} {\bibfield  {journal} {\bibinfo  {journal}
  {Phys. Rev. Lett.}\ }\textbf {\bibinfo {volume} {10}},\ \bibinfo {pages} {96}
  (\bibinfo {year} {1963})}\BibitemShut {NoStop}%
\bibitem [{\citenamefont {Yu}(1965)}]{Yu1965}%
  \BibitemOpen
  \bibfield  {author} {\bibinfo {author} {\bibfnamefont {L.}~\bibnamefont
  {Yu}},\ }\href {\doibase 10.7498/aps.21.75} {\bibfield  {journal} {\bibinfo
  {journal} {Acta Phys. Sin.}\ }\textbf {\bibinfo {volume} {21}},\ \bibinfo
  {pages} {75} (\bibinfo {year} {1965})}\BibitemShut {NoStop}%
\bibitem [{\citenamefont {Shiba}(1968)}]{Shiba1968}%
  \BibitemOpen
  \bibfield  {author} {\bibinfo {author} {\bibfnamefont {H.}~\bibnamefont
  {Shiba}},\ }\href@noop {} {\bibfield  {journal} {\bibinfo  {journal} {Prog.
  Theor. Phys.}\ }\textbf {\bibinfo {volume} {40}},\ \bibinfo {pages} {435}
  (\bibinfo {year} {1968})}\BibitemShut {NoStop}%
\bibitem [{\citenamefont {Rusinov}(1969)}]{Rusinov1969}%
  \BibitemOpen
  \bibfield  {author} {\bibinfo {author} {\bibfnamefont {A.~I.}\ \bibnamefont
  {Rusinov}},\ }\href@noop {} {\bibfield  {journal} {\bibinfo  {journal} {JETP
  Lett.}\ }\textbf {\bibinfo {volume} {9}},\ \bibinfo {pages} {85} (\bibinfo
  {year} {1969})}\BibitemShut {NoStop}%
\bibitem [{\citenamefont {Ji}\ \emph {et~al.}(2008)\citenamefont {Ji},
  \citenamefont {Zhang}, \citenamefont {Fu}, \citenamefont {Chen},
  \citenamefont {Ma}, \citenamefont {Li}, \citenamefont {Duan}, \citenamefont
  {Jia},\ and\ \citenamefont {Xue}}]{Ji2008}%
  \BibitemOpen
  \bibfield  {author} {\bibinfo {author} {\bibfnamefont {S.~H.}\ \bibnamefont
  {Ji}}, \bibinfo {author} {\bibfnamefont {T.}~\bibnamefont {Zhang}}, \bibinfo
  {author} {\bibfnamefont {Y.~S.}\ \bibnamefont {Fu}}, \bibinfo {author}
  {\bibfnamefont {X.}~\bibnamefont {Chen}}, \bibinfo {author} {\bibfnamefont
  {X.~C.}\ \bibnamefont {Ma}}, \bibinfo {author} {\bibfnamefont
  {J.}~\bibnamefont {Li}}, \bibinfo {author} {\bibfnamefont {W.~H.}\
  \bibnamefont {Duan}}, \bibinfo {author} {\bibfnamefont {J.~F.}\ \bibnamefont
  {Jia}}, \ and\ \bibinfo {author} {\bibfnamefont {Q.~K.}\ \bibnamefont
  {Xue}},\ }\href {\doibase 10.1103/PhysRevLett.100.226801} {\bibfield
  {journal} {\bibinfo  {journal} {Phys. Rev. Lett.}\ }\textbf {\bibinfo
  {volume} {100}},\ \bibinfo {pages} {226801} (\bibinfo {year}
  {2008})}\BibitemShut {NoStop}%
\bibitem [{\citenamefont {Hatter}\ \emph {et~al.}(2015)\citenamefont {Hatter},
  \citenamefont {Heinrich}, \citenamefont {Ruby}, \citenamefont {Pascual},\
  and\ \citenamefont {Franke}}]{Hatter2015}%
  \BibitemOpen
  \bibfield  {author} {\bibinfo {author} {\bibfnamefont {N.}~\bibnamefont
  {Hatter}}, \bibinfo {author} {\bibfnamefont {B.~W.}\ \bibnamefont
  {Heinrich}}, \bibinfo {author} {\bibfnamefont {M.}~\bibnamefont {Ruby}},
  \bibinfo {author} {\bibfnamefont {J.~I.}\ \bibnamefont {Pascual}}, \ and\
  \bibinfo {author} {\bibfnamefont {K.~J.}\ \bibnamefont {Franke}},\ }\href
  {\doibase 10.1038/ncomms9988} {\bibfield  {journal} {\bibinfo  {journal}
  {Nat. Commun.}\ }\textbf {\bibinfo {volume} {6}},\ \bibinfo {pages} {1}
  (\bibinfo {year} {2015})}\BibitemShut {NoStop}%
\bibitem [{\citenamefont {Ruby}\ \emph {et~al.}(2016)\citenamefont {Ruby},
  \citenamefont {Peng}, \citenamefont {{Von Oppen}}, \citenamefont {Heinrich},\
  and\ \citenamefont {Franke}}]{Ruby2016}%
  \BibitemOpen
  \bibfield  {author} {\bibinfo {author} {\bibfnamefont {M.}~\bibnamefont
  {Ruby}}, \bibinfo {author} {\bibfnamefont {Y.}~\bibnamefont {Peng}}, \bibinfo
  {author} {\bibfnamefont {F.}~\bibnamefont {{Von Oppen}}}, \bibinfo {author}
  {\bibfnamefont {B.~W.}\ \bibnamefont {Heinrich}}, \ and\ \bibinfo {author}
  {\bibfnamefont {K.~J.}\ \bibnamefont {Franke}},\ }\href {\doibase
  10.1103/PhysRevLett.117.186801} {\bibfield  {journal} {\bibinfo  {journal}
  {Phys. Rev. Lett.}\ }\textbf {\bibinfo {volume} {117}},\ \bibinfo {pages}
  {186801} (\bibinfo {year} {2016})}\BibitemShut {NoStop}%
\bibitem [{\citenamefont {Choi}\ \emph {et~al.}(2017)\citenamefont {Choi},
  \citenamefont {Rubio-Verd{\'{u}}}, \citenamefont {{De Bruijckere}},
  \citenamefont {Ugeda}, \citenamefont {Lorente},\ and\ \citenamefont
  {Pascual}}]{Choi2017}%
  \BibitemOpen
  \bibfield  {author} {\bibinfo {author} {\bibfnamefont {D.~J.}\ \bibnamefont
  {Choi}}, \bibinfo {author} {\bibfnamefont {C.}~\bibnamefont
  {Rubio-Verd{\'{u}}}}, \bibinfo {author} {\bibfnamefont {J.}~\bibnamefont {{De
  Bruijckere}}}, \bibinfo {author} {\bibfnamefont {M.~M.}\ \bibnamefont
  {Ugeda}}, \bibinfo {author} {\bibfnamefont {N.}~\bibnamefont {Lorente}}, \
  and\ \bibinfo {author} {\bibfnamefont {J.~I.}\ \bibnamefont {Pascual}},\
  }\href {\doibase 10.1038/ncomms15175} {\bibfield  {journal} {\bibinfo
  {journal} {Nat. Commun.}\ }\textbf {\bibinfo {volume} {8}},\ \bibinfo {pages}
  {15175} (\bibinfo {year} {2017})}\BibitemShut {NoStop}%
\bibitem [{\citenamefont {Liebhaber}\ \emph {et~al.}(2020)\citenamefont
  {Liebhaber}, \citenamefont {{Acero Gonz{\'{a}}lez}}, \citenamefont {Baba},
  \citenamefont {Reecht}, \citenamefont {Heinrich}, \citenamefont {Rohlf},
  \citenamefont {Rossnagel}, \citenamefont {{Von Oppen}},\ and\ \citenamefont
  {Franke}}]{Liebhaber2020}%
  \BibitemOpen
  \bibfield  {author} {\bibinfo {author} {\bibfnamefont {E.}~\bibnamefont
  {Liebhaber}}, \bibinfo {author} {\bibfnamefont {S.}~\bibnamefont {{Acero
  Gonz{\'{a}}lez}}}, \bibinfo {author} {\bibfnamefont {R.}~\bibnamefont
  {Baba}}, \bibinfo {author} {\bibfnamefont {G.}~\bibnamefont {Reecht}},
  \bibinfo {author} {\bibfnamefont {B.~W.}\ \bibnamefont {Heinrich}}, \bibinfo
  {author} {\bibfnamefont {S.}~\bibnamefont {Rohlf}}, \bibinfo {author}
  {\bibfnamefont {K.}~\bibnamefont {Rossnagel}}, \bibinfo {author}
  {\bibfnamefont {F.}~\bibnamefont {{Von Oppen}}}, \ and\ \bibinfo {author}
  {\bibfnamefont {K.~J.}\ \bibnamefont {Franke}},\ }\href {\doibase
  10.1021/acs.nanolett.9b03988} {\bibfield  {journal} {\bibinfo  {journal}
  {Nano Lett.}\ }\textbf {\bibinfo {volume} {20}},\ \bibinfo {pages} {339}
  (\bibinfo {year} {2020})},\ \Eprint {http://arxiv.org/abs/1903.09663}
  {arXiv:1903.09663} \BibitemShut {NoStop}%
\bibitem [{\citenamefont {Senkpiel}\ \emph {et~al.}(2019)\citenamefont
  {Senkpiel}, \citenamefont {Rubio-Verd{\'{u}}}, \citenamefont {Etzkorn},
  \citenamefont {Drost}, \citenamefont {Schoop}, \citenamefont {Dambach},
  \citenamefont {Padurariu}, \citenamefont {Kubala}, \citenamefont {Ankerhold},
  \citenamefont {Ast},\ and\ \citenamefont {Kern}}]{Senkpiel2019}%
  \BibitemOpen
  \bibfield  {author} {\bibinfo {author} {\bibfnamefont {J.}~\bibnamefont
  {Senkpiel}}, \bibinfo {author} {\bibfnamefont {C.}~\bibnamefont
  {Rubio-Verd{\'{u}}}}, \bibinfo {author} {\bibfnamefont {M.}~\bibnamefont
  {Etzkorn}}, \bibinfo {author} {\bibfnamefont {R.}~\bibnamefont {Drost}},
  \bibinfo {author} {\bibfnamefont {L.~M.}\ \bibnamefont {Schoop}}, \bibinfo
  {author} {\bibfnamefont {S.}~\bibnamefont {Dambach}}, \bibinfo {author}
  {\bibfnamefont {C.}~\bibnamefont {Padurariu}}, \bibinfo {author}
  {\bibfnamefont {B.}~\bibnamefont {Kubala}}, \bibinfo {author} {\bibfnamefont
  {J.}~\bibnamefont {Ankerhold}}, \bibinfo {author} {\bibfnamefont {C.~R.}\
  \bibnamefont {Ast}}, \ and\ \bibinfo {author} {\bibfnamefont
  {K.}~\bibnamefont {Kern}},\ }\href {\doibase 10.1103/PhysRevB.100.014502}
  {\bibfield  {journal} {\bibinfo  {journal} {Phys. Rev. B}\ }\textbf {\bibinfo
  {volume} {100}},\ \bibinfo {pages} {014502} (\bibinfo {year} {2019})},\
  \Eprint {http://arxiv.org/abs/1803.08726} {arXiv:1803.08726} \BibitemShut
  {NoStop}%
\bibitem [{\citenamefont {Choi}\ \emph {et~al.}(2018)\citenamefont {Choi},
  \citenamefont {Fern{\'{a}}ndez}, \citenamefont {Herrera}, \citenamefont
  {Rubio-Verd{\'{u}}}, \citenamefont {Ugeda}, \citenamefont {Guillam{\'{o}}n},
  \citenamefont {Suderow}, \citenamefont {Pascual},\ and\ \citenamefont
  {Lorente}}]{Choi2018}%
  \BibitemOpen
  \bibfield  {author} {\bibinfo {author} {\bibfnamefont {D.~J.}\ \bibnamefont
  {Choi}}, \bibinfo {author} {\bibfnamefont {C.~G.}\ \bibnamefont
  {Fern{\'{a}}ndez}}, \bibinfo {author} {\bibfnamefont {E.}~\bibnamefont
  {Herrera}}, \bibinfo {author} {\bibfnamefont {C.}~\bibnamefont
  {Rubio-Verd{\'{u}}}}, \bibinfo {author} {\bibfnamefont {M.~M.}\ \bibnamefont
  {Ugeda}}, \bibinfo {author} {\bibfnamefont {I.}~\bibnamefont
  {Guillam{\'{o}}n}}, \bibinfo {author} {\bibfnamefont {H.}~\bibnamefont
  {Suderow}}, \bibinfo {author} {\bibfnamefont {J.~I.}\ \bibnamefont
  {Pascual}}, \ and\ \bibinfo {author} {\bibfnamefont {N.}~\bibnamefont
  {Lorente}},\ }\href {\doibase 10.1103/PhysRevLett.120.167001} {\bibfield
  {journal} {\bibinfo  {journal} {Phys. Rev. Lett.}\ }\textbf {\bibinfo
  {volume} {120}},\ \bibinfo {pages} {167001} (\bibinfo {year}
  {2018})}\BibitemShut {NoStop}%
\bibitem [{\citenamefont {Saunderson}\ \emph
  {et~al.}(2020{\natexlab{a}})\citenamefont {Saunderson}, \citenamefont
  {Annett}, \citenamefont {{\'{U}}jfalussy}, \citenamefont {Csire},\ and\
  \citenamefont {Gradhand}}]{Saunderson2020}%
  \BibitemOpen
  \bibfield  {author} {\bibinfo {author} {\bibfnamefont {T.~G.}\ \bibnamefont
  {Saunderson}}, \bibinfo {author} {\bibfnamefont {J.~F.}\ \bibnamefont
  {Annett}}, \bibinfo {author} {\bibfnamefont {B.}~\bibnamefont
  {{\'{U}}jfalussy}}, \bibinfo {author} {\bibfnamefont {G.}~\bibnamefont
  {Csire}}, \ and\ \bibinfo {author} {\bibfnamefont {M.}~\bibnamefont
  {Gradhand}},\ }\href@noop {} {\bibfield  {journal} {\bibinfo  {journal}
  {Phys. Rev. B}\ }\textbf {\bibinfo {volume} {101}},\ \bibinfo {pages}
  {064510} (\bibinfo {year} {2020}{\natexlab{a}})}\BibitemShut {NoStop}%
\bibitem [{\citenamefont {Saunderson}\ \emph
  {et~al.}(2020{\natexlab{b}})\citenamefont {Saunderson}, \citenamefont
  {Gy\"{o}rgyp\'{a}l}, \citenamefont {Annett}, \citenamefont {Csire},
  \citenamefont {{\'{U}}jfalussy},\ and\ \citenamefont
  {Gradhand}}]{Saunderson2020_a}%
  \BibitemOpen
  \bibfield  {author} {\bibinfo {author} {\bibfnamefont {T.~G.}\ \bibnamefont
  {Saunderson}}, \bibinfo {author} {\bibfnamefont {Z.}~\bibnamefont
  {Gy\"{o}rgyp\'{a}l}}, \bibinfo {author} {\bibfnamefont {J.~F.}\ \bibnamefont
  {Annett}}, \bibinfo {author} {\bibfnamefont {G.}~\bibnamefont {Csire}},
  \bibinfo {author} {\bibfnamefont {B.}~\bibnamefont {{\'{U}}jfalussy}}, \ and\
  \bibinfo {author} {\bibfnamefont {M.}~\bibnamefont {Gradhand}},\ }\href@noop
  {} {\bibfield  {journal} {\bibinfo  {journal} {Phys. Rev. B}\ }\textbf
  {\bibinfo {volume} {102}},\ \bibinfo {pages} {245106} (\bibinfo {year}
  {2020}{\natexlab{b}})}\BibitemShut {NoStop}%
\bibitem [{\citenamefont {Oliveira}\ \emph {et~al.}(1988)\citenamefont
  {Oliveira}, \citenamefont {Gross},\ and\ \citenamefont
  {Kohn}}]{Oliveira1988}%
  \BibitemOpen
  \bibfield  {author} {\bibinfo {author} {\bibfnamefont {L.~N.}\ \bibnamefont
  {Oliveira}}, \bibinfo {author} {\bibfnamefont {E.~K.~U.}\ \bibnamefont
  {Gross}}, \ and\ \bibinfo {author} {\bibfnamefont {W.}~\bibnamefont {Kohn}},\
  }\href {\doibase 10.1103/PhysRevLett.60.2430} {\bibfield  {journal} {\bibinfo
   {journal} {Phys. Rev. Lett.}\ }\textbf {\bibinfo {volume} {60}},\ \bibinfo
  {pages} {2430} (\bibinfo {year} {1988})}\BibitemShut {NoStop}%
\bibitem [{\citenamefont {Dreizler}\ and\ \citenamefont
  {Gross}(1990)}]{DFT1990}%
  \BibitemOpen
  \bibfield  {author} {\bibinfo {author} {\bibfnamefont {R.~M.}\ \bibnamefont
  {Dreizler}}\ and\ \bibinfo {author} {\bibfnamefont {E.~K.~U.}\ \bibnamefont
  {Gross}},\ }\href@noop {} {\emph {\bibinfo {title} {{Density Functional
  Theory}}}}\ (\bibinfo  {publisher} {Springer, Berlin},\ \bibinfo {year}
  {1990})\BibitemShut {NoStop}%
\bibitem [{\citenamefont {Suvasani}\ \emph {et~al.}(1993)\citenamefont
  {Suvasani}, \citenamefont {Temmerman},\ and\ \citenamefont
  {Gy\"orffy}}]{Suvasani1993}%
  \BibitemOpen
  \bibfield  {author} {\bibinfo {author} {\bibfnamefont {M.~B.}\ \bibnamefont
  {Suvasani}}, \bibinfo {author} {\bibfnamefont {W.~M.}\ \bibnamefont
  {Temmerman}}, \ and\ \bibinfo {author} {\bibfnamefont {B.}~\bibnamefont
  {Gy\"orffy}},\ }\href@noop {} {\bibfield  {journal} {\bibinfo  {journal}
  {Phys. Rev. B}\ }\textbf {\bibinfo {volume} {48}},\ \bibinfo {pages} {1202}
  (\bibinfo {year} {1993})}\BibitemShut {NoStop}%
\bibitem [{\citenamefont {Csire}\ \emph {et~al.}(2015)\citenamefont {Csire},
  \citenamefont {{\'{U}}jfalussy}, \citenamefont {Cserti},\ and\ \citenamefont
  {Gy\"orffy}}]{Csire2015a}%
  \BibitemOpen
  \bibfield  {author} {\bibinfo {author} {\bibfnamefont {G.}~\bibnamefont
  {Csire}}, \bibinfo {author} {\bibfnamefont {B.}~\bibnamefont
  {{\'{U}}jfalussy}}, \bibinfo {author} {\bibfnamefont {J.}~\bibnamefont
  {Cserti}}, \ and\ \bibinfo {author} {\bibfnamefont {B.}~\bibnamefont
  {Gy\"orffy}},\ }\href {\doibase 10.1103/PhysRevB.91.165142} {\bibfield
  {journal} {\bibinfo  {journal} {Phys. Rev. B}\ }\textbf {\bibinfo {volume}
  {91}},\ \bibinfo {pages} {165142} (\bibinfo {year} {2015})}\BibitemShut
  {NoStop}%
\bibitem [{\citenamefont {Csire}\ \emph
  {et~al.}(2018{\natexlab{a}})\citenamefont {Csire}, \citenamefont
  {De{\'{a}}k}, \citenamefont {Ny{\'{a}}ri}, \citenamefont {Ebert},
  \citenamefont {Annett},\ and\ \citenamefont {{\'{U}}jfalussy}}]{Csire2018}%
  \BibitemOpen
  \bibfield  {author} {\bibinfo {author} {\bibfnamefont {G.}~\bibnamefont
  {Csire}}, \bibinfo {author} {\bibfnamefont {A.}~\bibnamefont {De{\'{a}}k}},
  \bibinfo {author} {\bibfnamefont {B.}~\bibnamefont {Ny{\'{a}}ri}}, \bibinfo
  {author} {\bibfnamefont {H.}~\bibnamefont {Ebert}}, \bibinfo {author}
  {\bibfnamefont {J.~F.}\ \bibnamefont {Annett}}, \ and\ \bibinfo {author}
  {\bibfnamefont {B.}~\bibnamefont {{\'{U}}jfalussy}},\ }\href {\doibase
  10.1103/PhysRevB.97.024514} {\bibfield  {journal} {\bibinfo  {journal} {Phys.
  Rev. B}\ }\textbf {\bibinfo {volume} {97}},\ \bibinfo {pages} {024514}
  (\bibinfo {year} {2018}{\natexlab{a}})}\BibitemShut {NoStop}%
\bibitem [{\citenamefont {Csire}\ \emph
  {et~al.}(2018{\natexlab{b}})\citenamefont {Csire}, \citenamefont
  {\'Ujfalussy},\ and\ \citenamefont {Annett}}]{Csire2018a}%
  \BibitemOpen
  \bibfield  {author} {\bibinfo {author} {\bibfnamefont {G.}~\bibnamefont
  {Csire}}, \bibinfo {author} {\bibfnamefont {B.}~\bibnamefont {\'Ujfalussy}},
  \ and\ \bibinfo {author} {\bibfnamefont {J.~F.}\ \bibnamefont {Annett}},\
  }\href@noop {} {\bibfield  {journal} {\bibinfo  {journal} {Eur. Phys. J. B}\
  }\textbf {\bibinfo {volume} {91}},\ \bibinfo {pages} {217} (\bibinfo {year}
  {2018}{\natexlab{b}})}\BibitemShut {NoStop}%
\bibitem [{\citenamefont {Ghosh}\ \emph {et~al.}(2020)\citenamefont {Ghosh},
  \citenamefont {Csire}, \citenamefont {Whittlesea}, \citenamefont {Annett},
  \citenamefont {Gradhand}, \citenamefont {\'Ujfalussy},\ and\ \citenamefont
  {Quintanilla}}]{Ghosh2020}%
  \BibitemOpen
  \bibfield  {author} {\bibinfo {author} {\bibfnamefont {S.~K.}\ \bibnamefont
  {Ghosh}}, \bibinfo {author} {\bibfnamefont {G.}~\bibnamefont {Csire}},
  \bibinfo {author} {\bibfnamefont {P.}~\bibnamefont {Whittlesea}}, \bibinfo
  {author} {\bibfnamefont {J.~F.}\ \bibnamefont {Annett}}, \bibinfo {author}
  {\bibfnamefont {M.}~\bibnamefont {Gradhand}}, \bibinfo {author}
  {\bibfnamefont {B.}~\bibnamefont {\'Ujfalussy}}, \ and\ \bibinfo {author}
  {\bibfnamefont {J.}~\bibnamefont {Quintanilla}},\ }\href {\doibase
  10.1103/PhysRevB.101.100506} {\bibfield  {journal} {\bibinfo  {journal}
  {Phys. Rev. B}\ }\textbf {\bibinfo {volume} {101}},\ \bibinfo {pages}
  {100506} (\bibinfo {year} {2020})}\BibitemShut {NoStop}%
\bibitem [{\citenamefont {Ruby}\ \emph {et~al.}(2015)\citenamefont {Ruby},
  \citenamefont {Heinrich}, \citenamefont {Pascual},\ and\ \citenamefont
  {Franke}}]{Ruby2015b}%
  \BibitemOpen
  \bibfield  {author} {\bibinfo {author} {\bibfnamefont {M.}~\bibnamefont
  {Ruby}}, \bibinfo {author} {\bibfnamefont {B.~W.}\ \bibnamefont {Heinrich}},
  \bibinfo {author} {\bibfnamefont {J.~I.}\ \bibnamefont {Pascual}}, \ and\
  \bibinfo {author} {\bibfnamefont {K.~J.}\ \bibnamefont {Franke}},\ }\href
  {\doibase 10.1103/PhysRevLett.114.157001} {\bibfield  {journal} {\bibinfo
  {journal} {Phys. Rev. Lett.}\ }\textbf {\bibinfo {volume} {114}},\ \bibinfo
  {pages} {157001} (\bibinfo {year} {2015})}\BibitemShut {NoStop}%
\bibitem [{\citenamefont {Dresselhaus}\ \emph {et~al.}(2007)\citenamefont
  {Dresselhaus}, \citenamefont {Dresselhaus},\ and\ \citenamefont
  {Jorio}}]{Dresselhaus2007}%
  \BibitemOpen
  \bibfield  {author} {\bibinfo {author} {\bibfnamefont {M.~S.}\ \bibnamefont
  {Dresselhaus}}, \bibinfo {author} {\bibfnamefont {G.}~\bibnamefont
  {Dresselhaus}}, \ and\ \bibinfo {author} {\bibfnamefont {A.}~\bibnamefont
  {Jorio}},\ }\href@noop {} {\emph {\bibinfo {title} {{Group Theory:
  Application to the Physics of Condensed Matter}}}}\ (\bibinfo  {publisher}
  {Springer},\ \bibinfo {year} {2007})\BibitemShut {NoStop}%
\bibitem [{\citenamefont {Balatsky}\ \emph {et~al.}(2006)\citenamefont
  {Balatsky}, \citenamefont {Vekhter},\ and\ \citenamefont
  {Zhu}}]{Balatsky2006}%
  \BibitemOpen
  \bibfield  {author} {\bibinfo {author} {\bibfnamefont {A.~V.}\ \bibnamefont
  {Balatsky}}, \bibinfo {author} {\bibfnamefont {I.}~\bibnamefont {Vekhter}}, \
  and\ \bibinfo {author} {\bibfnamefont {J.~X.}\ \bibnamefont {Zhu}},\ }\href
  {\doibase 10.1103/RevModPhys.78.373} {\bibfield  {journal} {\bibinfo
  {journal} {Rev. Mod. Phys.}\ }\textbf {\bibinfo {volume} {78}},\ \bibinfo
  {pages} {373} (\bibinfo {year} {2006})}\BibitemShut {NoStop}%
\bibitem [{\citenamefont {Schrieffer}(1967)}]{Schrieffer1967}%
  \BibitemOpen
  \bibfield  {author} {\bibinfo {author} {\bibfnamefont {J.~R.}\ \bibnamefont
  {Schrieffer}},\ }\href {\doibase 10.1063/1.1709517} {\bibfield  {journal}
  {\bibinfo  {journal} {J. Appl. Phys.}\ }\textbf {\bibinfo {volume} {38}},\
  \bibinfo {pages} {1143} (\bibinfo {year} {1967})}\BibitemShut {NoStop}%
\bibitem [{\citenamefont {Ruby}\ \emph {et~al.}(2018)\citenamefont {Ruby},
  \citenamefont {Heinrich}, \citenamefont {Peng}, \citenamefont {{Von Oppen}},\
  and\ \citenamefont {Franke}}]{Ruby2018}%
  \BibitemOpen
  \bibfield  {author} {\bibinfo {author} {\bibfnamefont {M.}~\bibnamefont
  {Ruby}}, \bibinfo {author} {\bibfnamefont {B.~W.}\ \bibnamefont {Heinrich}},
  \bibinfo {author} {\bibfnamefont {Y.}~\bibnamefont {Peng}}, \bibinfo {author}
  {\bibfnamefont {F.}~\bibnamefont {{Von Oppen}}}, \ and\ \bibinfo {author}
  {\bibfnamefont {K.~J.}\ \bibnamefont {Franke}},\ }\href {\doibase
  10.1103/PhysRevLett.120.156803} {\bibfield  {journal} {\bibinfo  {journal}
  {Phys. Rev. Lett.}\ }\textbf {\bibinfo {volume} {120}},\ \bibinfo {pages}
  {156803} (\bibinfo {year} {2018})}\BibitemShut {NoStop}%
\bibitem [{\citenamefont {K\"uster}\ \emph {et~al.}(2021)\citenamefont
  {K\"uster}, \citenamefont {Montero}, \citenamefont {Guimar\={a}es},
  \citenamefont {Lounis}, \citenamefont {Parkin},\ and\ \citenamefont
  {Sessi}}]{Kuster2021}%
  \BibitemOpen
  \bibfield  {author} {\bibinfo {author} {\bibfnamefont {F.}~\bibnamefont
  {K\"uster}}, \bibinfo {author} {\bibfnamefont {A.~M.}\ \bibnamefont
  {Montero}}, \bibinfo {author} {\bibfnamefont {S.}~\bibnamefont
  {Guimar\={a}es}, \bibfnamefont {Filipe S. M.and~Brinker}}, \bibinfo {author}
  {\bibfnamefont {S.}~\bibnamefont {Lounis}}, \bibinfo {author} {\bibfnamefont
  {S.~S.~P.}\ \bibnamefont {Parkin}}, \ and\ \bibinfo {author} {\bibfnamefont
  {P.}~\bibnamefont {Sessi}},\ }\href@noop {} {\bibfield  {journal} {\bibinfo
  {journal} {Nature Communications}\ }\textbf {\bibinfo {volume} {12}},\
  \bibinfo {pages} {1108} (\bibinfo {year} {2021})}\BibitemShut {NoStop}%
\bibitem [{\citenamefont {Zhang}\ \emph {et~al.}(2020)\citenamefont {Zhang},
  \citenamefont {Samuely}, \citenamefont {Iwahara}, , \citenamefont
  {Ka\v{c}mar\v{c}\'ik}, \citenamefont {Wang}, \citenamefont {May},
  \citenamefont {Jochum}, \citenamefont {Onufriienko}, \citenamefont {Szab\'o},
  \citenamefont {Zhou}, \citenamefont {Samuely}, \citenamefont {Moshchalkov},
  \citenamefont {Chibotaru},\ and\ \citenamefont {Rubahn}}]{Zhang2020}%
  \BibitemOpen
  \bibfield  {author} {\bibinfo {author} {\bibfnamefont {G.}~\bibnamefont
  {Zhang}}, \bibinfo {author} {\bibfnamefont {T.}~\bibnamefont {Samuely}},
  \bibinfo {author} {\bibfnamefont {N.}~\bibnamefont {Iwahara}}, , \bibinfo
  {author} {\bibfnamefont {J.}~\bibnamefont {Ka\v{c}mar\v{c}\'ik}}, \bibinfo
  {author} {\bibfnamefont {C.}~\bibnamefont {Wang}}, \bibinfo {author}
  {\bibfnamefont {P.~W.}\ \bibnamefont {May}}, \bibinfo {author} {\bibfnamefont
  {J.~K.}\ \bibnamefont {Jochum}}, \bibinfo {author} {\bibfnamefont
  {O.}~\bibnamefont {Onufriienko}}, \bibinfo {author} {\bibfnamefont
  {P.}~\bibnamefont {Szab\'o}}, \bibinfo {author} {\bibfnamefont
  {S.}~\bibnamefont {Zhou}}, \bibinfo {author} {\bibfnamefont {P.}~\bibnamefont
  {Samuely}}, \bibinfo {author} {\bibfnamefont {V.~V.}\ \bibnamefont
  {Moshchalkov}}, \bibinfo {author} {\bibfnamefont {L.~F.}\ \bibnamefont
  {Chibotaru}}, \ and\ \bibinfo {author} {\bibfnamefont {H.-G.}\ \bibnamefont
  {Rubahn}},\ }\href@noop {} {\bibfield  {journal} {\bibinfo  {journal}
  {Science Advances}\ }\textbf {\bibinfo {volume} {6}},\ \bibinfo {pages}
  {eaaz2536} (\bibinfo {year} {2020})}\BibitemShut {NoStop}%
\bibitem [{\citenamefont {Ruby}\ \emph {et~al.}(2017)\citenamefont {Ruby},
  \citenamefont {Heinrich}, \citenamefont {Peng}, \citenamefont {von Oppen},\
  and\ \citenamefont {Franke}}]{Ruby2017}%
  \BibitemOpen
  \bibfield  {author} {\bibinfo {author} {\bibfnamefont {M.}~\bibnamefont
  {Ruby}}, \bibinfo {author} {\bibfnamefont {B.~W.}\ \bibnamefont {Heinrich}},
  \bibinfo {author} {\bibfnamefont {Y.}~\bibnamefont {Peng}}, \bibinfo {author}
  {\bibfnamefont {F.}~\bibnamefont {von Oppen}}, \ and\ \bibinfo {author}
  {\bibfnamefont {K.~J.}\ \bibnamefont {Franke}},\ }\href {\doibase
  10.1021/acs.nanolett.7b01728} {\bibfield  {journal} {\bibinfo  {journal}
  {Nano Letters}\ }\textbf {\bibinfo {volume} {17}},\ \bibinfo {pages} {4473}
  (\bibinfo {year} {2017})},\ \bibinfo {note} {pMID: 28640633},\ \Eprint
  {http://arxiv.org/abs/https://doi.org/10.1021/acs.nanolett.7b01728}
  {https://doi.org/10.1021/acs.nanolett.7b01728} \BibitemShut {NoStop}%
\bibitem [{\citenamefont {Martin}\ \emph {et~al.}(1999)\citenamefont {Martin},
  \citenamefont {Litak}, \citenamefont {Gy\"orffy}, \citenamefont {Annett},\
  and\ \citenamefont {Wysoki\ifmmode~\acute{n}\else
  \'{n}\fi{}ski}}]{Martin1999}%
  \BibitemOpen
  \bibfield  {author} {\bibinfo {author} {\bibfnamefont {A.~M.}\ \bibnamefont
  {Martin}}, \bibinfo {author} {\bibfnamefont {G.}~\bibnamefont {Litak}},
  \bibinfo {author} {\bibfnamefont {B.~L.}\ \bibnamefont {Gy\"orffy}}, \bibinfo
  {author} {\bibfnamefont {J.~F.}\ \bibnamefont {Annett}}, \ and\ \bibinfo
  {author} {\bibfnamefont {K.~I.}\ \bibnamefont {Wysoki\ifmmode~\acute{n}\else
  \'{n}\fi{}ski}},\ }\href {\doibase 10.1103/PhysRevB.60.7523} {\bibfield
  {journal} {\bibinfo  {journal} {Phys. Rev. B}\ }\textbf {\bibinfo {volume}
  {60}},\ \bibinfo {pages} {7523} (\bibinfo {year} {1999})}\BibitemShut
  {NoStop}%
\bibitem [{\citenamefont {Moradian}\ \emph
  {et~al.}(2000{\natexlab{a}})\citenamefont {Moradian}, \citenamefont
  {Annett},\ and\ \citenamefont {Gy\"orffy}}]{Moradian2000}%
  \BibitemOpen
  \bibfield  {author} {\bibinfo {author} {\bibfnamefont {R.}~\bibnamefont
  {Moradian}}, \bibinfo {author} {\bibfnamefont {J.~F.}\ \bibnamefont
  {Annett}}, \ and\ \bibinfo {author} {\bibfnamefont {B.~L.}\ \bibnamefont
  {Gy\"orffy}},\ }\href {\doibase 10.1103/PhysRevB.62.3508} {\bibfield
  {journal} {\bibinfo  {journal} {Phys. Rev. B}\ }\textbf {\bibinfo {volume}
  {62}},\ \bibinfo {pages} {3508} (\bibinfo {year}
  {2000}{\natexlab{a}})}\BibitemShut {NoStop}%
\bibitem [{\citenamefont {Moradian}\ \emph
  {et~al.}(2000{\natexlab{b}})\citenamefont {Moradian}, \citenamefont {Annett},
  \citenamefont {Gy\"orffy},\ and\ \citenamefont {Litak}}]{Moradian2000a}%
  \BibitemOpen
  \bibfield  {author} {\bibinfo {author} {\bibfnamefont {R.}~\bibnamefont
  {Moradian}}, \bibinfo {author} {\bibfnamefont {J.~F.}\ \bibnamefont
  {Annett}}, \bibinfo {author} {\bibfnamefont {B.~L.}\ \bibnamefont
  {Gy\"orffy}}, \ and\ \bibinfo {author} {\bibfnamefont {G.}~\bibnamefont
  {Litak}},\ }\href {\doibase 10.1103/PhysRevB.63.024501} {\bibfield  {journal}
  {\bibinfo  {journal} {Phys. Rev. B}\ }\textbf {\bibinfo {volume} {63}},\
  \bibinfo {pages} {024501} (\bibinfo {year} {2000}{\natexlab{b}})}\BibitemShut
  {NoStop}%
\bibitem [{\citenamefont {Nadj-Perge}\ \emph {et~al.}(2013)\citenamefont
  {Nadj-Perge}, \citenamefont {Drozdov}, \citenamefont {Bernevig},\ and\
  \citenamefont {Yazdani}}]{Nadj-Perge2013}%
  \BibitemOpen
  \bibfield  {author} {\bibinfo {author} {\bibfnamefont {S.}~\bibnamefont
  {Nadj-Perge}}, \bibinfo {author} {\bibfnamefont {I.~K.}\ \bibnamefont
  {Drozdov}}, \bibinfo {author} {\bibfnamefont {B.~A.}\ \bibnamefont
  {Bernevig}}, \ and\ \bibinfo {author} {\bibfnamefont {A.}~\bibnamefont
  {Yazdani}},\ }\href {\doibase 10.1103/PhysRevB.88.020407} {\bibfield
  {journal} {\bibinfo  {journal} {Phys. Rev. B}\ }\textbf {\bibinfo {volume}
  {88}},\ \bibinfo {pages} {020407} (\bibinfo {year} {2013})}\BibitemShut
  {NoStop}%
\bibitem [{\citenamefont {R\"ontynen}\ and\ \citenamefont
  {Ojanen}(2015)}]{TopoSup2015}%
  \BibitemOpen
  \bibfield  {author} {\bibinfo {author} {\bibfnamefont {J.}~\bibnamefont
  {R\"ontynen}}\ and\ \bibinfo {author} {\bibfnamefont {T.}~\bibnamefont
  {Ojanen}},\ }\href {\doibase 10.1103/PhysRevLett.114.236803} {\bibfield
  {journal} {\bibinfo  {journal} {Phys. Rev. Lett.}\ }\textbf {\bibinfo
  {volume} {114}},\ \bibinfo {pages} {236803} (\bibinfo {year}
  {2015})}\BibitemShut {NoStop}%
\bibitem [{\citenamefont {Kim}\ \emph {et~al.}(2018)\citenamefont {Kim},
  \citenamefont {Palacio-Morales}, \citenamefont {Posske}, \citenamefont
  {R{\'{o}}zsa}, \citenamefont {Palot{\'{a}}s}, \citenamefont {Szunyogh},
  \citenamefont {Thorwart},\ and\ \citenamefont {Wiesendanger}}]{Kim2018}%
  \BibitemOpen
  \bibfield  {author} {\bibinfo {author} {\bibfnamefont {H.}~\bibnamefont
  {Kim}}, \bibinfo {author} {\bibfnamefont {A.}~\bibnamefont
  {Palacio-Morales}}, \bibinfo {author} {\bibfnamefont {T.}~\bibnamefont
  {Posske}}, \bibinfo {author} {\bibfnamefont {L.}~\bibnamefont {R{\'{o}}zsa}},
  \bibinfo {author} {\bibfnamefont {K.}~\bibnamefont {Palot{\'{a}}s}}, \bibinfo
  {author} {\bibfnamefont {L.}~\bibnamefont {Szunyogh}}, \bibinfo {author}
  {\bibfnamefont {M.}~\bibnamefont {Thorwart}}, \ and\ \bibinfo {author}
  {\bibfnamefont {R.}~\bibnamefont {Wiesendanger}},\ }\href {\doibase
  10.1126/sciadv.aar5251} {\bibfield  {journal} {\bibinfo  {journal} {Sci.
  Adv.}\ }\textbf {\bibinfo {volume} {4}},\ \bibinfo {pages} {1} (\bibinfo
  {year} {2018})}\BibitemShut {NoStop}%
\bibitem [{\citenamefont {Schneider}\ \emph {et~al.}(2020)\citenamefont
  {Schneider}, \citenamefont {Brinker}, \citenamefont {Steinbrecher},
  \citenamefont {Hermenau}, \citenamefont {Posske}, \citenamefont {Dias},
  \citenamefont {Lounis}, \citenamefont {Wiesendanger},\ and\ \citenamefont
  {Wiebe}}]{Schneider2020}%
  \BibitemOpen
  \bibfield  {author} {\bibinfo {author} {\bibfnamefont {L.}~\bibnamefont
  {Schneider}}, \bibinfo {author} {\bibfnamefont {S.}~\bibnamefont {Brinker}},
  \bibinfo {author} {\bibfnamefont {M.}~\bibnamefont {Steinbrecher}}, \bibinfo
  {author} {\bibfnamefont {J.}~\bibnamefont {Hermenau}}, \bibinfo {author}
  {\bibfnamefont {T.}~\bibnamefont {Posske}}, \bibinfo {author} {\bibfnamefont
  {M.~d.~S.}\ \bibnamefont {Dias}}, \bibinfo {author} {\bibfnamefont
  {S.}~\bibnamefont {Lounis}}, \bibinfo {author} {\bibfnamefont
  {R.}~\bibnamefont {Wiesendanger}}, \ and\ \bibinfo {author} {\bibfnamefont
  {J.}~\bibnamefont {Wiebe}},\ }\href
  {https://doi.org/10.1038/s41467-020-18540-3} {\bibfield  {journal} {\bibinfo
  {journal} {Nature Communications}\ }\textbf {\bibinfo {volume} {11}},\
  \bibinfo {pages} {4707} (\bibinfo {year} {2020})}\BibitemShut {NoStop}%
\end{thebibliography}
\end{document}